\documentclass[aip,reprint]{revtex4-1}
\usepackage{graphicx}
\usepackage{dcolumn}
\usepackage{bm}
\usepackage{amsmath}%
\usepackage{amssymb}

\begin{document}

\title{Carrier-envelope phase sensitive inversion in two-level systems}
\author{Christian Jirauschek}
\affiliation{Department of Electrical Engineering and Computer Science
and Research Laboratory of Electronics, Massachusetts Institute of Technology,
77 Massachusetts Avenue, Cambridge, Massachusetts 02139}
\author{Lingze Duan}
\affiliation{Department of Electrical Engineering and Computer Science
and Research Laboratory of Electronics, Massachusetts Institute of Technology,
77 Massachusetts Avenue, Cambridge, Massachusetts 02139}
\author{Oliver D. M\"{u}cke}
\affiliation{Department of Electrical Engineering and Computer Science
and Research Laboratory of Electronics, Massachusetts Institute of Technology,
77 Massachusetts Avenue, Cambridge, Massachusetts 02139}
\author{Franz X. K\"{a}rtner}
\affiliation{Department of Electrical Engineering and Computer Science
and Research Laboratory of Electronics, Massachusetts Institute of Technology,
77 Massachusetts Avenue, Cambridge, Massachusetts 02139}
\author{Martin Wegener}%
\affiliation{Institut f\"{u}r Angewandte Physik, Universit\"{a}t Karlsruhe (TH),
Wolfgang-Gaede-Stra\ss e 1, 76131 Karlsruhe, Germany}%
\author{Uwe Morgner}
\affiliation{Max-Planck-Institut f\"{u}r Kernphysik, Saupfercheckweg 1, 69117 Heidelberg, Germany}%
\date{17 June 2011, published as J. Opt. Soc. Am. B 22, 2065--2075 (2005)}

\begin{abstract}
We theoretically study the carrier-envelope phase dependent inversion
generated in a two-level system by excitation with a few-cycle pulse. Based
on the invariance of the inversion under time reversal of the exciting
field, parameters are introduced to characterize the phase sensitivity of
the induced inversion. Linear and nonlinear phase effects are numerically
studied for rectangular and sinc-shaped pulses. Furthermore, analytical
results are obtained in the limits of weak fields as well as strong
dephasing, and by nearly degenerate perturbation theory for sinusoidal
excitation. The results show that the phase sensitive inversion in the ideal
two-level system is a promising route for constructing carrier-envelope
phase detectors.
\newline \copyright 2005 Optical Society of America
\end{abstract}

\maketitle %% NULL FUNCTION WITH LATEX 2e; required for REVTeX4
\section{Introduction}

The two-level system is a fundamental and widely used model to describe the
interaction of electromagnetic waves with matter. The temporal evolution of
the system is usually treated within the framework of the rotating wave
approximation (RWA), where the driving electric field enters the equations
of motion only through its complex envelope and center frequency.\cite{all75}
For a complete description of the electric field, also the carrier-envelope
(CE) phase has to be taken into account, specifying the position of the
envelope with respect to the rapidly oscillating carrier wave. The RWA
breaks down for strongly driven systems,\cite{hug98} giving rise to new
effects which not only depend on the pulse envelope and carrier frequency,
but also\ on the CE phase.

The phase sensitive dynamics of the driven two-level system beyond the RWA
has been the topic of several papers. The phase dependence of the inversion
was carefully examined for a sinusoidal excitation, serving as a model for
the interaction of atoms and molecules with continuous wave laser radiation.%
\cite{shi65,gus72,ahm75,mol76,mol78,gri98} The discussion was extended to\
rectangular pulses, obtained by abruptly switching on and off the sinusoidal
excitation. The phase sensitivity of the inversion was investigated and
experimentally demonstrated in the radio frequency regime by exciting the
anticrossing of the potassium 21s - 19f states.\cite{gri98}

Following the arrival of laser pulses consisting of only a few optical
cycles, there has been considerable interest in phase sensitive effects in
the pulsed optical regime \cite{xu96}, and various approaches have been used
for detecting the CE phase/frequency.\cite%
{apo00,jon00,mor01,muc02B,for04,apo04,pau03} In this context, the CE phase
dependent emission of two-level systems\cite{iva93} and semiconductors\cite%
{muc02} interacting with few-cycle pulses has been theoretically
investigated, and the effect has experimentally been observed in GaAs.\cite%
{muc04} So far, less attention has been given to the CE phase dependence of
the inversion, with a few exceptions studying the interaction with Gaussian
pulses.\cite{tho85,bro98}

While for sinusoidal excitation, the generated inversion shows a CE phase
dependence\ even for weak fields, the phase sensitive dynamics relies
completely on nonlinear effects for pulsed optical excitation. In this
paper, both linear and higher order phase dependent inversion effects in
two-level systems are\ theoretically investigated, using analytical
approximations and numerical simulations with properly chosen test pulses.
In addition, the influence of dephasing on the phase sensitivity is studied.
General properties of the steady state inversion are discussed, and
approximate expressions are derived in the linear and the nonlinear regime.
The paper is organized as follows: In Section 2, the equations of motion for
an excited two-level system are given in a fixed and a rotating reference
frame. In Section 3, general properties of the\ steady state inversion are
discussed. In Section 4, the phase dependent inversion in the weak field
limit is analytically examined, and in Section 5, the discussion is extended
beyond the linear regime for rectangular and sinc-shaped pulses, using
numerical simulations and nearly degenerate perturbation theory. In Section
6, the influence of the phase relaxation is studied based on the strong
dephasing approximation and numerical results. We conclude in Section 7.

\section{Equations of motion}

A two-level system is characterized by its dipole matrix element $d$ and
resonance frequency $\omega _{ba}=2\pi f_{ba}=(E_{b}-E_{a})/\hbar $, where $%
E_{a}$ and $E_{b}$ are the eigenenergies associated with the low and high
energy states $\left\vert a\right\rangle $ and $\left\vert b\right\rangle $,
respectively. Dissipative effects, which arise due to the interaction of the
ideal two-level system with its environment, can be taken into account in a
statistical approach. The density matrix is represented by the components of
the Bloch vector $\mathbf{s}$. The components $s_{1}$ and $s_{2}$ are
related to the real and imaginary parts of the off-diagonal density matrix
elements by $\rho _{ab}=\left( s_{1}+\mathrm{i}s_{2}\right) /2$, and $\rho
_{bb}-\rho _{aa}=s_{3}=w$ is the population inversion. In this paper, the
relaxation processes are described by phenomenological parameters, the
energy relaxation rate $\gamma _{1}=1/T_{1}$ and the dephasing rate $\gamma
_{2}=1/T_{2}$. Frequently, relaxation is dominated by processes which lead
to a destruction of the phase coherence in the quantum system without
affecting the inversion, resulting in a dephasing time $T_{2}$ which is much
shorter than the energy relaxation time $T_{1}$. For example, this is
typically the case in a gas due to collision broadening.\cite{aku92} Thus,
in the following, we set $\gamma _{1}=0$, assuming that the energy
relaxation processes are slow compared to the interaction time with the
field, whereas we do allow for dephasing processes occurring on a time scale
comparable to the duration of the exciting pulse.

Assuming linear polarization of the exciting field and a vanishing static
dipole moment, the dynamics of the system is described by the Bloch
equations,\cite{all75}%
\begin{eqnarray}
\dot{s}_{1} &=&-\omega _{ba}s_{2}-\gamma _{2}s_{1},  \notag \\
\dot{s}_{2} &=&\omega _{ba}s_{1}+2\Omega s_{3}-\gamma _{2}s_{2},  \label{blo}
\\
\dot{s}_{3} &=&-2\Omega s_{2}.  \notag
\end{eqnarray}
The overdot denotes a time derivative. The electric field $E\left( t\right) $
is parametrized in terms of the instantaneous Rabi frequency $\Omega \left(
t\right) =dE\left( t\right) /\hbar $, which can be written as%
\begin{eqnarray}
\Omega \left( t\right)  &=&\Omega _{\mathrm{R}}\big[\varepsilon \left(
t\right) \exp \left( -\mathrm{i}\omega _{\mathrm{c}}t+\mathrm{i}\phi _{%
\mathrm{CE}}\right)   \notag \\
&&+\varepsilon ^{\ast }\left( t\right) \exp \left( \mathrm{i}\omega _{%
\mathrm{c}}t-\mathrm{i}\phi _{\mathrm{CE}}\right) \big]/2,  \label{Om}
\end{eqnarray}%
where the asterisk denotes the complex conjugate. Here, $\varepsilon \left(
t\right) $ is the normalized, in general complex envelope function, $\omega
_{\mathrm{c}}=2\pi f_{c}$ is the carrier frequency, and $\phi _{\mathrm{CE}}$
is the CE phase. In this paper, we refer to $\Omega _{\mathrm{R}%
}=dE_{0}/\hbar $ as the (peak) Rabi frequency, with the maximum value of the
electric field envelope $E_{0}$.

For some applications, it is convenient to transform the Bloch equations,
Eq.\ (\ref{blo}), into a rotating reference frame. The transformation
between the Bloch vector $\mathbf{s}$ and the vector in the rotating frame $%
\mathbf{u=}\left( u,v,w\right) ^{\mathrm{T}}$, where $\mathrm{T}$ denotes
the transposed vector, is given by $\mathbf{u}\left( t\right) =A\left(
t\right) \mathbf{s}\left( t\right) $, 
\begin{equation}
A\left( t\right) =\left( 
\begin{array}{ccc}
\cos \left( \omega _{ba}t\right) & \sin \left( \omega _{ba}t\right) & 0 \\ 
-\sin \left( \omega _{ba}t\right) & \cos \left( \omega _{ba}t\right) & 0 \\ 
0 & 0 & 1%
\end{array}%
\right) .
\end{equation}%
For $\gamma _{1}=0$, we obtain the equations of motion in the rotating frame%
\begin{subequations}%
\label{Bloch_rf}%
\begin{eqnarray}
\dot{u} &=&2\Omega \sin \left( \omega _{ba}t\right) w-\gamma _{2}u,
\label{Bloch_rf_a} \\
\dot{v} &=&2\Omega \cos \left( \omega _{ba}t\right) w-\gamma _{2}v,
\label{Bloch_rf_b} \\
\dot{w} &=&-2\Omega \sin \left( \omega _{ba}t\right) u-2\Omega \cos \left(
\omega _{ba}t\right) v.  \label{Bloch_rf_c}
\end{eqnarray}%
\end{subequations}%

\section{Steady state inversion}

We consider a two-level system excited by an electric field pulse,
parametrized in terms of $\Omega \left( t\right) $, and neglect energy
relaxation, i.e., $\gamma _{1}=0$. During interaction with the field, the
inversion in the system evolves and reaches a steady state value $w_{\mathrm{%
s}}$, staying constant after interaction with the pulse. In this paper, we
investigate the CE phase sensitivity of the steady state inversion, which is
readily accessible to experiments. We assume that the dipole moment of the
system is initially vanishing, corresponding to the initial condition $%
s_{1}=s_{2}=0$ and $s_{3}=w_{0}$. This is always fulfilled if the system
starts from equilibrium, where $w_{0}$ corresponds to the equilibrium
inversion.

\subsection{Inversion invariance under time reversal of the field}

Strong excitation of the two-level system results in the breakdown of the
RWA, and of related features like the area theorem.\cite{hug98} By way of
contrast, the property discussed in the following is even valid in the
strong field limit.

\begin{figure}[th]
\centering\includegraphics{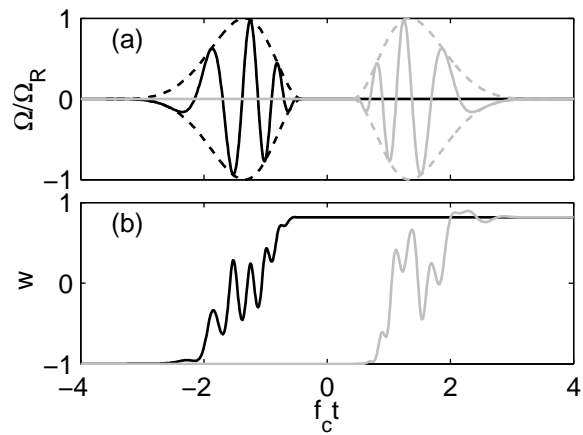}
\caption{Dynamics of the two-level system for excitation with a field $%
\Omega \left( t\right) $ (black curve) and the corresponding time-reversed
field $\Omega \left( -t\right) $ (gray curve). (a) Time dependent electric
field, parametrized in terms of the normalized Rabi frequency. (b) Evolution
of the inversion for a Rabi frequency $\Omega _{\mathrm{R}}=1.3\,\protect%
\omega _{\mathrm{c}}$, a transition frequency $\protect\omega _{ba}=1.7\,%
\protect\omega _{\mathrm{c}}$, and a dephasing rate $\protect\gamma _{2}=f_{%
\mathrm{c}}/10$.}
\label{cep3_1}
\end{figure}

\begin{figure}[th]
\centering\includegraphics{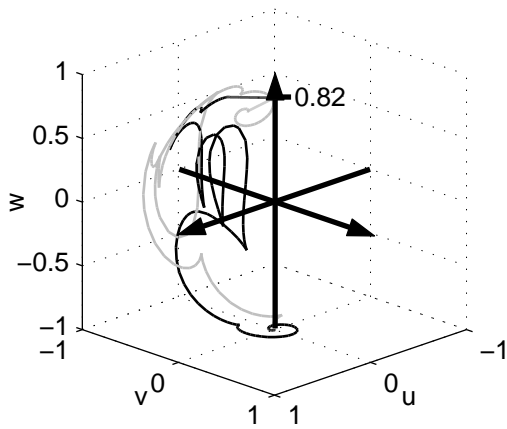}
\caption{Bloch vector trajectories for excitation of a two-level system with
a driving field (black curve) and the time-reversed field (gray curve). The
fields and system parameters are the same as for Fig.\ \ref{cep3_1}.}
\label{cep3_2}
\end{figure}

We take the Bloch equations, Eq. (\ref{Bloch_rf}), assuming $\gamma _{1}=0$
and an initially vanishing dipole moment. Under these conditions, the steady
state inversion $w_{\mathrm{s}}$ after interaction with an arbitrary pulse $%
\Omega \left( t\right) $ is the same as for excitation with the
time-reversed pulse $\Omega \left( -t\right) $. The proof is given in
Appendix A. This feature is remarkable, since the temporal evolution of the
system from its initial state is completely different in both cases, and
should not be confused with the time-reversed dynamics of the system. In
Fig.\ \ref{cep3_1}, the invariance of $w_{\mathrm{s}}$ with respect to time
reversal of the field is illustrated by means of an example. The time
dependent field of an exciting pulse is shown in Fig.\ \ref{cep3_1}(a), as
well as the corresponding time-reversed field. The pulse shape is in
principle arbitrary; the only condition is that the pulse has a finite
energy, and thus the field approaches zero for $t\rightarrow \pm \infty $.
For a propagating electric field, we furthermore require a vanishing dc
component. This condition, however, is not crucial for the inversion
invariance discussed here. In Fig.\ \ref{cep3_1}(b), the time dependent
inversion is displayed, starting at the initial value $w_{0}=-1$. The
evolution of the inversion is completely different for excitation with the
field and the time-reversed field, but still, the same steady state value is
reached in both cases. This can also be seen in the Bloch sphere
representation. In Fig.\ \ref{cep3_2}, the trajectories of the Bloch
components $u$, $v$ and $w$ in the rotating frame are shown, with the south
pole of the sphere corresponding to the initial state. The Bloch vectors
evolve in completely different ways, but the same steady state inversion is
reached. In the presence of dephasing, the case which is shown here, the
trajectories approach the same point for $t\rightarrow \infty $,\ since the
dipole moment decays, i.e., $u$ and $v$ tend to zero for $t\rightarrow
\infty $. By way of contrast, for $\gamma _{2}=0$, $u$ and $v$ approach
different steady state values for excitation with the field and the
time-reversed field, respectively. Thus, for $t\rightarrow \infty $, the two
trajectories still reach the same value for $w$, but not for $u$ and $v$.

\subsection{Steady state inversion for symmetric pulses}

In Sections 5 and 6, we discuss the two-level dynamics for excitation by
sinc-shaped and rectangular pulses, which are chirp-free and have symmetric
intensity envelopes. For such pulses (or more generally for pulses with $%
\varepsilon \left( t\right) =\varepsilon ^{\ast }\left( -t\right) $), the
time reversal of the field corresponds to a mere sign change of the CE
phase, as can be seen from Eq. (\ref{Om}). As a consequence of the time
reversal invariance discussed above, we have $w_{\mathrm{s}}\left( \phi _{%
\mathrm{CE}}\right) =w_{\mathrm{s}}\left( -\phi _{\mathrm{CE}}\right) $ for
these pulses. Due to the centrosymmetry of the two-level system, the
inversion is furthermore invariant with respect to a sign change of the
electric field, corresponding to a $\pi $ shift in the CE phase, i.e., $w_{%
\mathrm{s}}\left( \phi _{\mathrm{CE}}\right) =w_{\mathrm{s}}\left( \phi _{%
\mathrm{CE}}+\pi \right) $. Thus, the CE phase dependent occupation
probability can be represented by a $\pi $ periodic Fourier series. If $w_{%
\mathrm{s}}\left( \phi _{\mathrm{CE}}\right) =w_{\mathrm{s}}\left( -\phi _{%
\mathrm{CE}}\right) $ holds,\ only the even (i.e., $\cos $) terms remain.
For non-resonant excitation and non-excessive field strengths, the series
can be truncated after the first order term, and we obtain%
\begin{equation}
w_{\mathrm{s}}\left( \phi _{\mathrm{CE}}\right) =\bar{w}_{\mathrm{s}}+\Delta
\cos \left( 2\phi _{\mathrm{CE}}\right) .  \label{ws}
\end{equation}%
The parameters used here are the CE phase averaged inversion, 
\begin{equation}
\bar{w}_{\mathrm{s}}=\left[ w_{\mathrm{s}}\left( 0\right) +w_{\mathrm{s}%
}\left( \pi /2\right) \right] /2,
\end{equation}%
and the modulation amplitude%
\begin{equation}
\Delta =\left[ w_{\mathrm{s}}\left( 0\right) -w_{\mathrm{s}}\left( \pi
/2\right) \right] /2,
\end{equation}%
with $-1\leq \bar{w}_{\mathrm{s}},\Delta \leq 1$.

\begin{figure}[th]
\centering\includegraphics{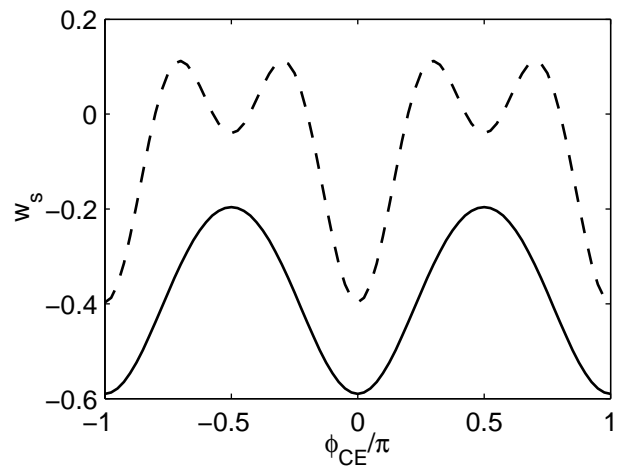}
\caption{Phase dependent steady state inversion $w_{\mathrm{s}}$ after
interaction with a two-cycle sinc pulse, for Rabi frequencies $\Omega _{%
\mathrm{R}}=1.5\,\protect\omega _{\mathrm{c}}$ (solid curve) and $\Omega _{%
\mathrm{R}}=2.3\,\protect\omega _{c}$ (dashed curve). The transition
frequency is $\protect\omega _{ba}=1.5\,\protect\omega _{\mathrm{c}}$, and
the dephasing time is $T_{2}=10/f_{\mathrm{c}}$.}
\label{cep3_3}
\end{figure}

Fig.\ \ref{cep3_3} shows the phase dependence of $w_{\mathrm{s}}$ after
interaction of the two-level system with a sinc pulse. For $\Omega _{\mathrm{%
R}}=1.5\,\omega _{\mathrm{c}}$, the phase dependence of $w_{\mathrm{s}}$ can
be well approximated by Eq.\ (\ref{ws}). For $\Omega _{\mathrm{R}%
}=2.3\,\omega _{\mathrm{c}}$, higher order terms in the Fourier expansion
result in a bump around the symmetry point. For excitation near the
two-level resonances $\omega _{ba}=\left( 2n+1\right) \omega _{\mathrm{c}}$, 
$n=0,1,2,\ldots $, the approximation Eq.\ (\ref{ws}) already breaks down for
smaller Rabi frequencies, i.e., more terms in the Fourier series are
necessary for a full description.

\section{Weak field limit}

As mentioned in Section 1, linear phase sensitive effects have been observed
for custom tailored microwave and radio frequency pulses, while for
laser-generated optical pulses, a CE phase dependence can only be obtained
in the nonlinear regime. The phase dependent dynamics of an excited
two-level system can be treated analytically by a time dependent
perturbation series expansion, with the first order approximation describing
the evolution in the linear regime. We use the Bloch equations in the
rotating reference frame, Eq. (\ref{Bloch_rf}). As initial condition at $%
t\rightarrow -\infty $, we choose $\mathbf{u}_{0}=\left( 0,0,w_{0}\right) ^{%
\mathrm{T}}$.\ By formally integrating Eqs. (\ref{Bloch_rf_a}) and (\ref%
{Bloch_rf_b}) and inserting the results into Eq. (\ref{Bloch_rf_c}), we
arrive at an integral equation for $w\left( t\right) $,%
\begin{eqnarray}
w\left( t\right)  &=&w_{0}-4\int_{-\infty }^{t}\mathrm{d}t^{\prime
}\int_{0}^{\infty }\mathrm{d}\tau \,\cos \left( \omega _{ba}\tau \right)  
\notag \\
&&\times \exp \left( -\gamma _{2}\tau \right) \Omega \left( t^{\prime
}\right) \Omega \left( t^{\prime }-\tau \right) w\left( t^{\prime }-\tau
\right) .  \label{pert}
\end{eqnarray}%
In the weak field limit, the time dependent inversion changes only slightly
with respect to its initial value. The first order approximation (which is
second order in $\Omega _{\mathrm{R}}$) is obtained by inserting the $0$th
order solution $w^{\left( 0\right) }\left( t\right) =w_{0}$ into the
right-hand side of Eq. (\ref{pert}),%
\begin{eqnarray}
w^{\left( 1\right) }\left( t\right)  &=&w_{0}\left[ 1-4\int_{0}^{\infty }%
\mathrm{d}\tau \,\cos \left( \omega _{ba}\tau \right) \exp \left( -\gamma
_{2}\tau \right) \right.   \notag \\
&&\times \left. \int_{-\infty }^{t}\mathrm{d}t^{\prime }\,\Omega \left(
t^{\prime }\right) \Omega \left( t^{\prime }-\tau \right) \right] .
\label{pert2}
\end{eqnarray}%
The result for the steady state inversion $w_{\mathrm{s}}=w\left(
t\rightarrow \infty \right) $ is favorably expressed in terms of the Fourier
transform%
\begin{equation}
\Omega \left( \omega \right) =\int_{-\infty }^{\infty }\mathrm{d}t\,\,\Omega
\left( t\right) \exp \left( \mathrm{i}\omega t\right) ,  \label{four}
\end{equation}%
yielding%
\begin{equation}
w_{\mathrm{s}}^{\left( 1\right) }=w_{0}\left[ 1-\int_{0}^{\infty }\mathrm{d}%
\omega \,\,\,\left| \Omega \left( \omega \right) \right| ^{2}H\left( \omega
\right) \right] .  \label{u3}
\end{equation}%
Thus, in the linear regime, the relationship between $w_{\mathrm{s}}$ and
the power spectrum of the pulse $\left| \Omega \left( \omega \right) \right|
^{2}$ can be described in terms of a spectral filter function $H\left(
\omega \right) $, which is given by%
\begin{equation}
H\left( \omega \right) =\frac{4}{\pi }\frac{\gamma _{2}\left( \omega
^{2}+\gamma _{2}^{2}+\omega _{ba}^{2}\right) }{\left( \omega ^{2}+\gamma
_{2}^{2}+\omega _{ba}^{2}\right) ^{2}-4\omega _{ba}^{2}\omega ^{2}}.
\label{H(w)}
\end{equation}%
For $\gamma _{2}\ll \omega _{ba}$, we can make the approximation%
\begin{equation}
H\left( \omega \right) =\frac{4}{\pi }\frac{\gamma _{2}/2}{\left( \omega
-\omega _{ba}\right) ^{2}+\gamma _{2}^{2}},
\end{equation}%
and in the limit of no dephasing, $\gamma _{2}\rightarrow 0$, we obtain%
\begin{equation}
H\left( \omega \right) =2\delta \left( \omega -\omega _{ba}\right) .
\label{H0}
\end{equation}

The Fourier transform of $\Omega \left( t\right) $, Eq. (\ref{four}), can
with Eq. (\ref{Om}) be written as 
\begin{eqnarray}
\Omega \left( \omega \right)  &=&\Omega _{\mathrm{R}}\big[\varepsilon \left(
\omega -\omega _{\mathrm{c}}\right) \exp \left( \mathrm{i}\phi _{\mathrm{CE}%
}\right)   \notag \\
&&+\varepsilon ^{\ast }\left( -\omega -\omega _{\mathrm{c}}\right) \exp
\left( -\mathrm{i}\phi _{\mathrm{CE}}\right) \big]/2.  \label{Om_w}
\end{eqnarray}%
The power spectrum $\left\vert \Omega \left( \omega \right) \right\vert ^{2}$
is only phase independent if $\varepsilon \left( \omega \right) =0$ for $%
\omega \leq -\omega _{\mathrm{c}}$, which, for chirp-free pulses,
corresponds to the condition $\varepsilon \left( \omega \right) =0$ for $%
\left\vert \omega \right\vert \geq \omega _{\mathrm{c}}$.\cite{mor01}

Formally, Eq. (\ref{pert}) is a Volterra integral equation of the second
kind, as can be shown by exchanging the order of integration. The weak field
approximation, Eq. (\ref{pert2}), is thus equivalent to a first order Neumann
series expansion,\cite{jer99} yielding $\left( w_{\mathrm{s}}-w_{0}\right)
\propto \Omega _{\mathrm{R}}^{2}$ and also $\Delta \propto \Omega _{\mathrm{R%
}}^{2}$ for pulses with phase dependent power spectra. Thus, the amount of
generated inversion and its phase dependence scale linearly with the peak
intensity of the pulse. This is different for pulses with phase independent
power spectra,\ where in the weak field limit, the phase dependence of $w_{%
\mathrm{s}}$ is provided by the second order term. In this case, we obtain $%
\Delta \propto \Omega _{\mathrm{R}}^{4}$ (or $\Delta \propto \Omega _{%
\mathrm{R}}^{3}$ for two-level systems with a static dipole moment\cite%
{bro98}), i.e., the generated inversion becomes phase insensitive for weak
fields.

In the microwave regime, pulses with the desired envelopes and CE phases can
be custom tailored. For example, pulses with durations far below one carrier
cycle are available, giving rise to linear CE phase dependent processes.\cite%
{mit02} For the experimental demonstration of the phase sensitive inversion,
pulses in the radio frequency regime with approximately rectangular
envelopes were applied,\cite{gri98} which have a phase dependent power
spectrum. For such pulses, linear filter effects can be used to determine
the CE phase, and a phase dependent inversion $w_{\mathrm{s}}$ is obtained
even in the weak field limit. Strictly speaking, this is also true for
Gaussian or hyperbolic secant pulse shapes, often used as a model for laser
pulses.\cite{tho85,bro98,mor99} In contrast, for optical pulses emitted by
current laser systems, the power spectrum is phase independent. As a
consequence, a CE phase dependent inversion $w_{\mathrm{s}}$ can only be
observed in the nonlinear regime.

\section{Systems without dissipation}

In the following, we investigate the phase dependence of the steady state
inversion for interaction with rectangular and sinc pulses, serving as
typical model pulses for excitation in the radio frequency and optical
regime, respectively. In this section, we neglect dissipation, i.e., we set $%
\gamma _{1}=\gamma _{2}=0$. The rectangular pulse corresponds to a
sinusoidal field of finite duration and is thus closely related to
continuous wave excitation, which plays an important role in atomic and
molecular physics and quantum chemistry.\cite%
{shi65,gus72,ahm75,mol76,mol78,gri98} This particularly basic pulse shape
features a piecewise constant envelope. The periodicity of the field can be
exploited for approximately solving the equations of motion, and basic
features of the solutions can be derived. In the radio frequency regime,
rectangular pulse shapes can be approximately generated by switching on and
off a sinusoidal field, and they were used for the experimental
demonstration of the phase sensitive inversion.\cite{gri98} As already
pointed out, rectangular pulses have a phase dependent power spectrum,
yielding phase sensitive effects even in the linear regime. Such pulses are
currently not available in the optical regime from laser sources. Sinc
pulses, on the other hand, feature a rectangular, phase independent power
spectrum, and have proven to be a fairly good description of few-cycle laser
pulses.\cite{mor99}

\subsection{Excitation with rectangular pulses}

\begin{figure}[th]
\centering\includegraphics{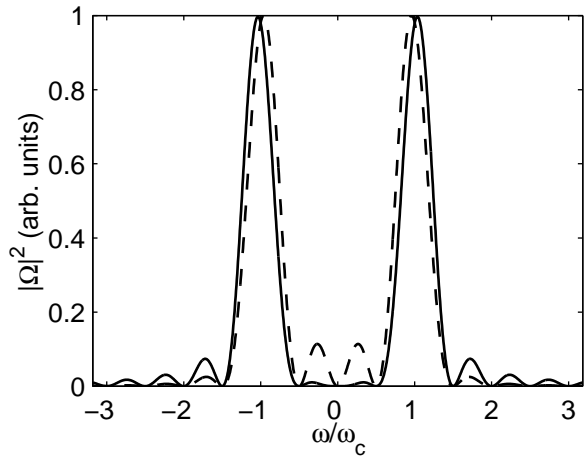}
\caption{Rectangular two-cycle pulse: Power spectrum for CE phases $0$
(solid line) and $\protect\pi /2$ (dashed line).}
\label{cep5_21}
\end{figure}

For rectangular pulses, the envelope in Eq. (\ref{Om}) is given by $%
\varepsilon \left( t\right) =1$ for $-T/2\leq t\leq T/2$, and $\varepsilon
\left( t\right) =0$ otherwise, where $T$ is the pulse duration. In order to
avoid a non-propagating dc component for arbitrary CE phases $\phi _{\mathrm{%
CE}}$, the pulses must contain an integer number of optical cycles, i.e., $T$
is an integer multiple of the carrier period $T_{0}=1/f_{\mathrm{c}}$. For $%
\phi _{\mathrm{CE}}\neq \left( n+1/2\right) \pi $, $n\in \mathbb{Z}$, the
field strength exhibits discontinuities, giving rise to a CE phase dependent
power spectrum. In Fig.\ \ref{cep5_21}, the power spectrum $\left\vert
\Omega \left( \omega \right) \right\vert ^{2}$ of a rectangular two-cycle
pulse is displayed. The CE phase dependence of the power spectrum becomes
also clear from the spectral properties of the pulse. Since the pulse
spectrum does not vanish identically for $\left\vert \omega \right\vert \geq
\omega _{\mathrm{c}}$, the two components in Eq. (\ref{Om_w}), centered
around $\omega _{\mathrm{c}}$ and $-\omega _{\mathrm{c}}$, have a region of
spectral overlap, resulting in a CE phase dependent interference.

\begin{figure*}[th]
\centering\includegraphics{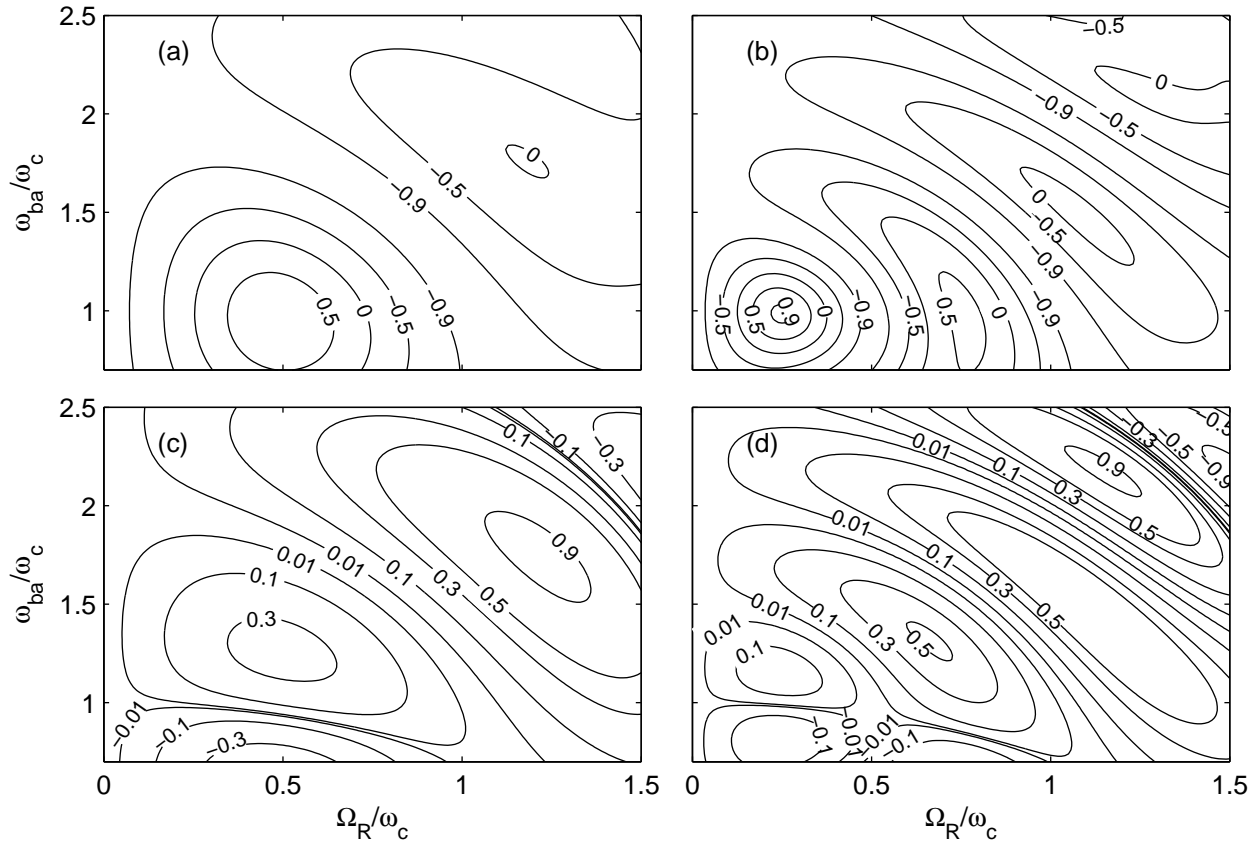}
\caption{Average inversion $\bar{w}_{\mathrm{s}}$ and modulation amplitude $%
\Delta $ as a function of the Rabi frequency $\Omega _{\mathrm{R}}$ and the
transition frequency $\protect\omega _{ba}$ in units of $\protect\omega _{%
\mathrm{c}}$. Displayed is $\bar{w}_{\mathrm{s}}$ after interaction with
rectangular (a) single-cycle and (b) two-cycle pulses, and $\Delta $ after
interaction with rectangular (c) single-cycle and (d) two-cycle pulses.}
\label{cep5_4}
\end{figure*}

In the following, we investigate the phase sensitivity of the inversion by
numerically solving the Bloch equations, Eq.\ (\ref{blo}). Fig.\ \ref{cep5_4}
shows the average inversion $\bar{w}_{\mathrm{s}}$ and the modulation
amplitude $\Delta $ as a function of $\Omega _{\mathrm{R}}$ and $\omega
_{ba} $. In Fig.\ \ref{cep5_4}(a) and (b), $\bar{w}_{\mathrm{s}}$ is
displayed for excitation with rectangular single-cycle and two-cycle pulses,
respectively. For weak fields, the inversion is largest for near-resonant
excitation. This resonance is more pronounced for the two-cycle pulses, due
to their narrower spectrum as compared to single-cycle pulses. For Rabi
frequencies approaching or even exceeding $\omega _{\mathrm{c}}$, regions
with strong inversion can also be found for off-resonant excitation due to
higher order transitions. $\Delta $ is shown in Fig.\ \ref{cep5_4}(c) and
(d), again for excitation with rectangular single-cycle and two-cycle
pulses. During interaction with the pulse, the inversion performs
oscillations due to carrier-wave Rabi flopping,\cite{hug98} and the steady
state value after interaction with the pulse $w_{\mathrm{s}}$ strongly
depends on the pulse and system parameters. As a consequence, for $\bar{w}_{%
\mathrm{s}}$ as well as $\Delta $, regions with positive and negative values
alternate. For rectangular pulses, the generated inversion exhibits a phase
dependence even in the linear regime, as discussed in Section 4. We
introduce the relative modulation amplitude or modulation depth%
\begin{equation}
\delta =\Delta /\left( \bar{w}_{\mathrm{s}}-w_{0}\right) ,  \label{delta}
\end{equation}%
which is helpful to discuss the phase sensitivity in the weak field limit. A
value of $\delta =\pm 1$ indicates maximum possible CE phase modulation of
the generated inversion, $\delta =0$ indicates no modulation. While the
amount of generated inversion vanishes for low field strengths, the relative
phase sensitivity of $w_{\mathrm{s}}$ does not disappear in the case of
phase dependent power spectra, yielding a finite value for $\delta $ in the
weak field limit.

\begin{figure}[th]
\centering\includegraphics{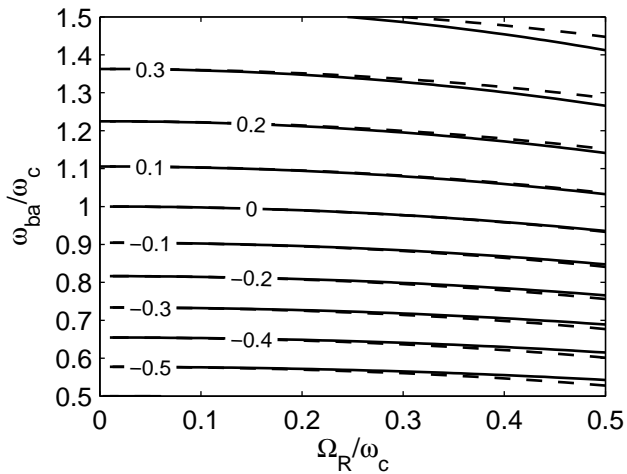}
\caption{Modulation depth $\protect\delta $ as a function of the Rabi
frequency $\Omega _{\mathrm{R}}$ and the transition frequency $\protect%
\omega _{ba}$ in units of $\protect\omega _{\mathrm{c}}$: Perturbative
(dashed curves) and exact numerical result (solid curves).}
\label{cep5_3}
\end{figure}

For interaction with a rectangular pulse shape, $\delta $ does not depend on
the pulse duration $T$. This is a consequence of the periodic excitation, as
shown in Appendix B, and is only true in two-level systems without
dissipation. An approximate solution beyond the weak field limit can be
obtained by using almost degenerate perturbation theory,\cite%
{cer70,hir78,ara84} where the perturbation parameter is the normalized Rabi
frequency $\Omega _{\mathrm{R}}/\omega _{\mathrm{c}}$. The zeroth order
perturbation theory, corresponding to the rotating wave approximation,\cite%
{ara84} does not contain any CE phase dependence and thus fails in
predicting phase sensitive effects.\cite{bro98} Following the procedure in
Ref.\ 28, we obtain the first order perturbative result%
\begin{equation}
\delta _{\mathrm{p}}=\frac{\left( \omega _{ba}^{2}-\omega _{\mathrm{c}%
}^{2}\right) \left( \omega _{ba}+\omega _{\mathrm{c}}\right) +\omega _{%
\mathrm{c}}\Omega _{\mathrm{R}}^{2}}{\left( \omega _{ba}^{2}+\omega _{%
\mathrm{c}}^{2}\right) \left( \omega _{ba}+\omega _{\mathrm{c}}\right)
-\omega _{\mathrm{c}}\Omega _{\mathrm{R}}^{2}},  \label{delta_p}
\end{equation}%
see Appendix C. Fig.\ \ref{cep5_3} shows the perturbative approximation and
the numerical solution for the modulation depth. The perturbative result is
in good agreement with the exact solution, especially for moderate field
strengths and moderate detuning, where the almost degenerate perturbative
treatment converges best.\cite{ara84} For small Rabi frequencies $\Omega _{%
\mathrm{R}}$, $w_{\mathrm{s}}$ approaches zero, but the relative phase
sensitivity of the inversion does not disappear, i.e., the modulation depth
stays finite. In the limit $\Omega _{\mathrm{R}}\rightarrow 0$, where the
perturbation theory becomes exact and coincides with the weak field
approximation, Eqs.\ (\ref{u3}) and (\ref{H0}), we obtain%
\begin{equation}
\delta _{\mathrm{p}}=\left( \omega _{ba}^{2}-\omega _{\mathrm{c}}^{2}\right)
/\left( \omega _{ba}^{2}+\omega _{\mathrm{c}}^{2}\right)   \label{delta_p3}
\end{equation}%
(except for odd multiphoton resonances, where $\delta $ becomes degenerate
for $\Omega _{\mathrm{R}}\rightarrow 0$ and the approximation breaks down).
Thus, in the weak field limit, $\delta >0$ for excitation below resonance,
and $\delta <0$ for excitation above resonance. For resonant excitation ($%
\omega _{\mathrm{c}}=\omega _{ba}$), where the population transfer reaches
its maximum, we obtain $\delta =0$, i.e., the inversion does not show a
phase sensitivity.

As already mentioned, the RWA does not predict any phase dependence in the
inversion. This might be\ counterintuitive, especially for pulses with CE
phase dependent power spectra, where the generated inversion depends on the
CE phase even in the weak field limit. A straightforward explanation is that
the validity of the RWA is restricted to near-resonant excitation,\cite%
{aku92} while on the other hand, the modulation depth in Eq. (\ref{delta_p3}%
) indicates a phase sensitivity of the inversion just for excitation off
resonance.

\subsection{Excitation with sinc-shaped pulses}

\begin{figure*}[th]
\centering\includegraphics{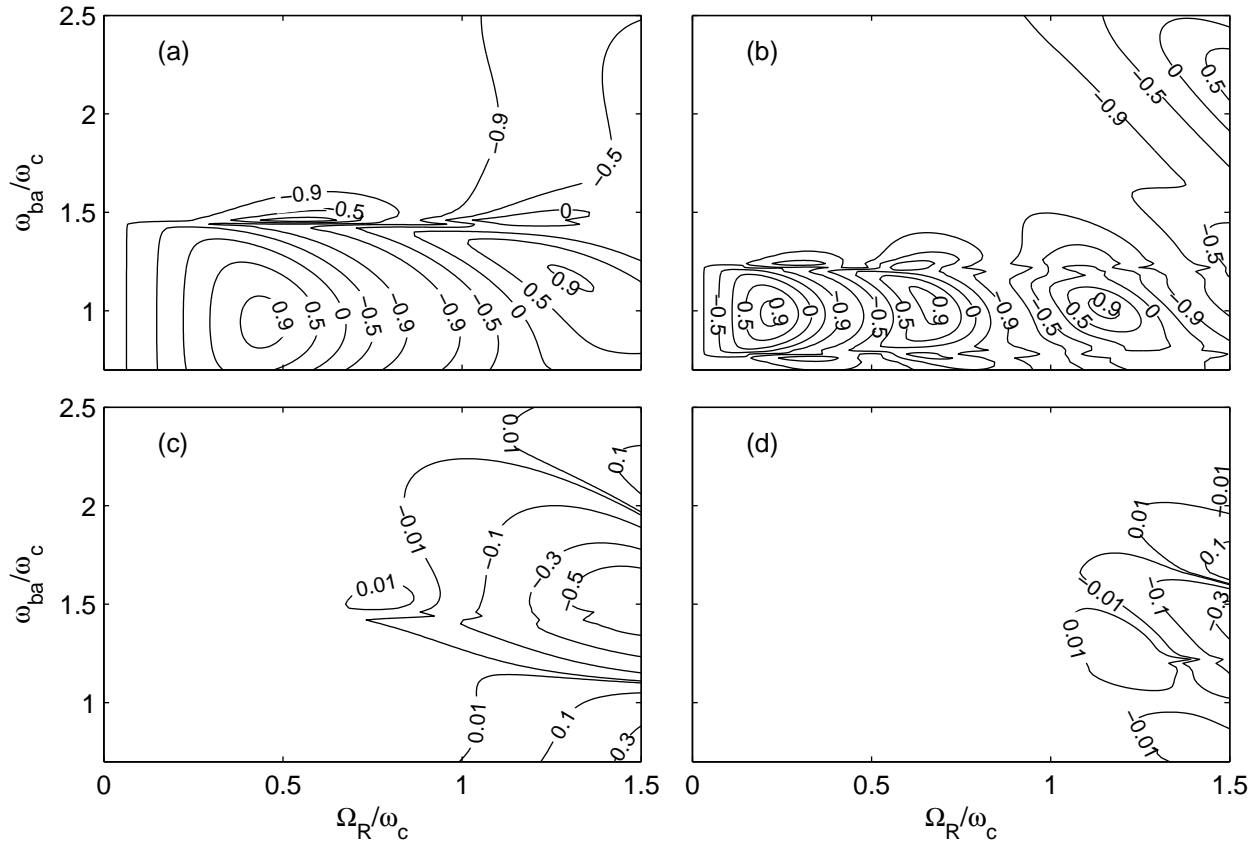}
\caption{Average inversion $\bar{w}_{\mathrm{s}}$ and modulation amplitude $%
\Delta $ as a function of the Rabi frequency $\Omega _{\mathrm{R}}$ and the
transition frequency $\protect\omega _{ba}$ in units of $\protect\omega _{%
\mathrm{c}}$. Displayed is $\bar{w}_{\mathrm{s}}$ after interaction with
sinc-shaped (a) single-cycle and (b) two-cycle pulses, and $\Delta $ after
interaction with sinc-shaped (c) single-cycle and (d) two-cycle pulses.}
\label{cep6_1}
\end{figure*}

Sinc pulses are characterized by the full width at half maximum (FWHM) value 
$T$ of the intensity envelope and the Rabi frequency $\Omega _{\mathrm{R}}$.
The normalized envelope function $\varepsilon \left( t\right) $, introduced
in Eq.\ (\ref{Om}), is given by $\varepsilon \left( t\right) =\mathrm{sinc}%
\left( t/\tau \right) =\sin \left( t/\tau \right) /\left( t/\tau \right) $.
The FWHM pulse duration is $T=2.783\tau $. These pulses have a rectangular
spectrum, extending from $f_{\mathrm{c}}-1/\left( 2\pi \tau \right) $ to $f_{%
\mathrm{c}}+1/\left( 2\pi \tau \right) $. We require $\tau >1/\left( 2\pi f_{%
\mathrm{c}}\right) $ in order to avoid unphysical dc components. Under this
condition, the power spectrum is CE phase independent, see discussion after
Eq.\ (\ref{Om_w}). Thus the inversion does not exhibit a CE phase
sensitivity in the linear regime. We note that Gaussian or hyperbolic secant
pulse descriptions feature unphysical dc components and a phase sensitive
power spectrum. This can lead to erroneous results, especially in the weak
field limit or for strong dephasing, see Section 6, where the remaining
phase sensitivity is due to the phase dependent power spectrum, as opposed
to nonlinear electric field effects.

The Bloch equations, Eq.\ (\ref{blo}), are numerically solved. In Fig.\ \ref%
{cep6_1}, the average inversion $\bar{w}_{\mathrm{s}}$ and the modulation
amplitude $\Delta $ are displayed as a function of $\Omega _{\mathrm{R}}$
and $\omega _{ba}$. In Fig.\ \ref{cep6_1}(a) and (b), $\bar{w}_{\mathrm{s}}$
is shown for excitation with sinc-shaped single-cycle and two-cycle pulses,
and the corresponding $\Delta $ is displayed in Fig.\ \ref{cep6_1}(c) and
(d). As in the case of excitation with rectangular pulses, see Fig.\ \ref%
{cep5_4}, regions with positive and negative values alternate for both $\bar{%
w}_{\mathrm{s}}$ and $\Delta $. The inversion $\bar{w}_{\mathrm{s}}$ is
largest for near-resonant excitation, and the resonance is more pronounced
for two-cycle pulses, which have a narrower spectrum. Higher order
transitions become relevant if $\Omega _{\mathrm{R}}$ approaches or even
exceeds $\omega _{\mathrm{c}}$, leading to regions with strong inversion
also for off-resonant excitation. As discussed in Section 4, the steady
state inversion shows no phase dependence in the linear regime, i.e., the
modulation depth $\delta $, Eq. (\ref{delta}), approaches zero for sinc
pulses. A significant phase sensitivity can only be found for field
strengths in the regime of carrier-wave Rabi flopping, where nonlinear
effects play a significant role. For sinc-shaped pulses, which have smooth
envelopes unlike rectangular pulses, the phase sensitivity of $w_{\mathrm{s}%
} $ depends considerably on the pulse duration, since the phase dependence
of the field is more pronounced for shorter pulses. Thus, higher field
strengths are necessary to obtain a significant modulation amplitude $\Delta 
$ for excitation by two-cycle pulses as compared to single-cycle pulses.

\section{Influence of dephasing}

In the following, we investigate the influence of phase relaxation processes
on the CE phase dependent inversion. In the limit of strong dephasing, an
approximate expression can be derived. We start from the integral equation
Eq.\ (\ref{pert}). Since for strong dephasing the Rabi oscillations of the
inversion are significantly dampened and the kernel decays rapidly, we can
approximate $w\left( t^{\prime }-\tau \right) \approx w\left( t^{\prime
}\right) $. By differentiation with respect to $t$, we obtain the first
order differential equation%
\begin{equation}
\dot{w}\!\left( t\right) =-4w\!\left( t\right)\!\int_{0}^{\infty }\!\!\!\!\mathrm{d}\tau
\,\cos\!\left( \omega _{ba}\tau \right) \exp\!\left( -\gamma _{2}\tau \right)
\Omega\!\left( t^{\prime }\right) \Omega\!\left( t^{\prime }-\tau \right) 
\label{deph1}
\end{equation}%
with the solution%
\begin{eqnarray}
w\left( t\right)  &=&w_{0}\exp \left[ -4\int_{0}^{\infty }\mathrm{d}\tau
\,\cos \left( \omega _{ba}\tau \right) \exp \left( -\gamma _{2}\tau \right)
\right.   \notag \\
&&\times \left. \int_{-\infty }^{t}\mathrm{d}t^{\prime }\,\Omega \left(
t^{\prime }\right) \Omega \left( t^{\prime }-\tau \right) \right] .
\label{deph2}
\end{eqnarray}%
For weak fields, where the exponential can be approximated by a first order
Taylor series, this expression coincides with the weak field approximation,
Eq.\ (\ref{pert2}). In analogy to Eq. (\ref{u3}), we can write the steady
state inversion as%
\begin{equation}
w_{\mathrm{s}}=w_{0}\exp \left[ -\int_{0}^{\infty }\mathrm{d}\omega \,\left|
\Omega \left( \omega \right) \right| ^{2}H\left( \omega \right) \right] ,
\label{deph3}
\end{equation}%
with the spectral filter function $H\left( \omega \right) $ defined in Eq.\ (%
\ref{H(w)}), and the Fourier transform $\Omega \left( \omega \right) $ given
in Eq.\ (\ref{four}). As discussed in Section 4, this expression contains a
CE phase sensitivity only for pulses with phase dependent power spectrum. If 
$\gamma _{2}$ is large compared to $\omega _{ba}$, $\omega _{\mathrm{c}}$
and the spectral width of the electric field, the steady state inversion is
given by%
\begin{eqnarray}
w_{\mathrm{s}} &=&w_{0}\exp \left[ -\frac{4}{\pi \gamma _{2}}%
\int_{0}^{\infty }\mathrm{d}\omega \,\left| \Omega \left( \omega \right)
\right| ^{2}\right]   \notag \\
&=&w_{0}\exp \left[ -\frac{4}{\gamma _{2}}\int_{-\infty }^{\infty }\mathrm{d}%
t\,\left| \Omega \left( t\right) \right| ^{2}\right] ,
\end{eqnarray}
which does not depend on the CE phase or the pulse shape, but only on the
total pulse energy. This result can also be obtained by adiabatic
elimination of the polarization in Eq. (\ref{Bloch_rf}).

\begin{figure}[th]
\centering\includegraphics{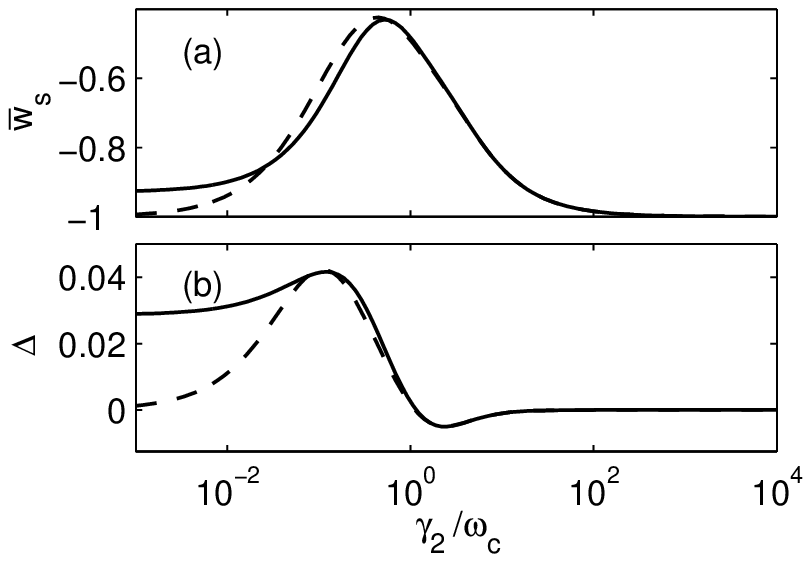}
\caption{Excitation with a rectangular two-cycle pulse: Exact numerical
result (solid curves) and strong dephasing approximation (dashed curves) for
the (a) average inversion $\bar{w}_{\mathrm{s}}$ and (b) modulation
amplitude $\Delta $ as a function of the normalized dephasing rate $\protect%
\gamma _{2}/\protect\omega _{\mathrm{c}}$. The Rabi frequency is $\Omega _{%
\mathrm{R}}=0.25\,\protect\omega _{\mathrm{c}}$, the transition frequency is 
$\protect\omega _{ba}=1.5\,\protect\omega _{\mathrm{c}}$.}
\label{cep6_2}
\end{figure}

\begin{figure}[th]
\centering\includegraphics{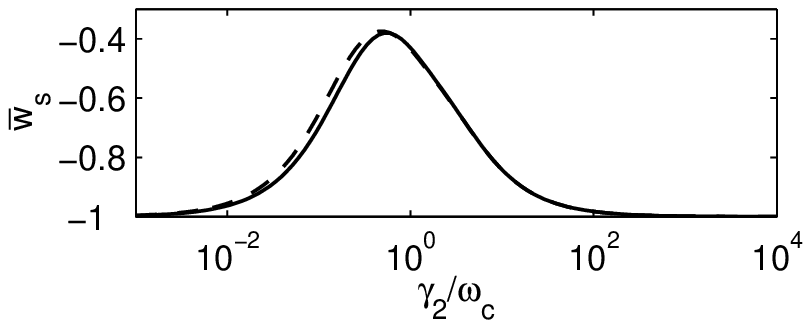}
\caption{Excitation with a sinc-shaped two-cycle pulse: Exact numerical
result (solid curve) and strong dephasing approximation (dashed curve) for
the average inversion $\bar{w}_{\mathrm{s}}$ as a function of the normalized
dephasing rate $\protect\gamma _{2}/\protect\omega _{\mathrm{c}}$. The Rabi
frequency is $\Omega _{\mathrm{R}}=0.25\,\protect\omega _{\mathrm{c}}$, the
transition frequency is $\protect\omega _{ba}=1.5\,\protect\omega _{\mathrm{c%
}}$.}
\label{cep6_3}
\end{figure}

\begin{figure}[th]
\centering\includegraphics{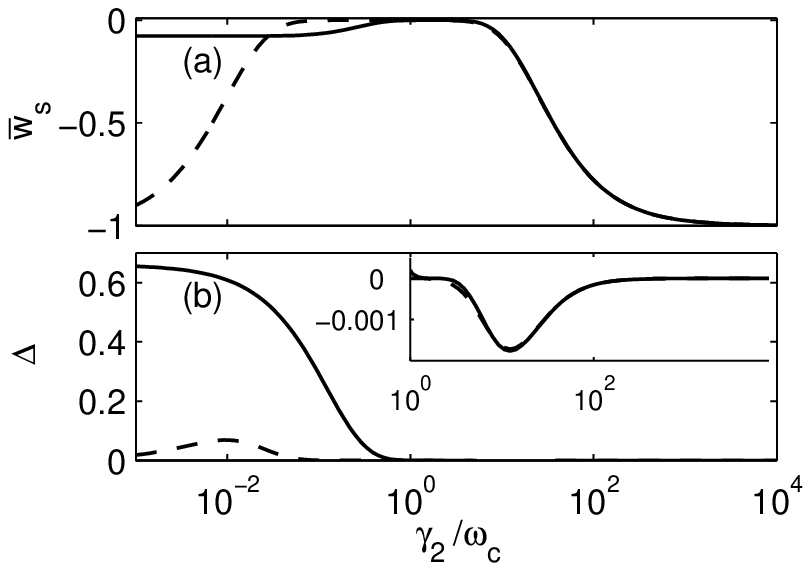}
\caption{Excitation with a rectangular two-cycle pulse: Exact numerical
result (solid curves) and strong dephasing approximation (dashed curves) for
the (a) average inversion $\bar{w}_{\mathrm{s}}$ and (b) modulation
amplitude $\Delta $ as a function of the normalized dephasing rate $\protect%
\gamma _{2}/\protect\omega _{\mathrm{c}}$. The Rabi frequency is $\Omega _{%
\mathrm{R}}=\protect\omega _{\mathrm{c}}$, the transition frequency is $%
\protect\omega _{ba}=1.5\,\protect\omega _{\mathrm{c}}$. The inset shows $%
\Delta $ at an increased scale.}
\label{cep6_4}
\end{figure}

\begin{figure}[th]
\centering\includegraphics{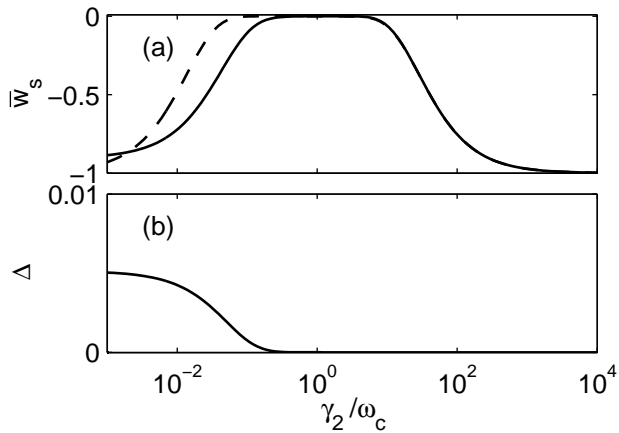}
\caption{Excitation with a sinc-shaped two-cycle pulse: Exact numerical
result (solid curves) and strong dephasing approximation (dashed curve) for
the (a) average inversion $\bar{w}_{\mathrm{s}}$ and (b) modulation
amplitude $\Delta $ as a function of the normalized dephasing rate $\protect%
\gamma _{2}/\protect\omega _{\mathrm{c}}$. The Rabi frequency is $\Omega _{%
\mathrm{R}}=\protect\omega _{\mathrm{c}}$, the transition frequency is $%
\protect\omega _{ba}=1.5\,\protect\omega _{\mathrm{c}}$. For $\Delta $, only
the numerical result is shown, since the approximation is not CE phase
sensitive for sinc pulses.}
\label{cep6_5}
\end{figure}

In Figs. \ref{cep6_2} - \ref{cep6_5}, the results of the numerical
simulation (solid curves) and the strong dephasing approximation (dashed
curves) for the CE phase dependent steady state inversion are shown as a
function of $\gamma _{2}/\omega _{\mathrm{c}}$. In Fig. \ref{cep6_2}, the
phase averaged inversion $\bar{w}_{\mathrm{s}}$ and the modulation amplitude 
$\Delta $ are displayed for interaction with a rectangular pulse, assuming a
moderate Rabi frequency $\Omega _{\mathrm{R}}=0.25\,\omega _{\mathrm{c}}$.
The results of the approximation and the numerical simulation agree well,
especially for increased dephasing, $\gamma _{2}\gtrsim 0.1\,\omega _{%
\mathrm{c}}$. Fig. \ref{cep6_3} shows the phase averaged inversion $\bar{w}_{%
\mathrm{s}}$ for excitation by sinc-shaped pulses with the same parameters
as before. Here, even better agreement between numerical and analytical
results is found. The CE phase dependence of the inversion is negligible in
this case, $\left\vert \Delta \right\vert <10^{-7}$, since nonlinear phase
sensitive effects do not play a significant role at those field strengths.
Figs. \ref{cep6_4} and \ref{cep6_5} show again the results for excitation by
rectangular and sinc-shaped pulses, respectively, but now for a higher Rabi
frequency $\Omega _{\mathrm{R}}=\omega _{\mathrm{c}}$. In this case, the
validity range of the strong dephasing approximation is reduced as compared
to $\Omega _{\mathrm{R}}=0.25\,\omega _{\mathrm{c}}$, and good agreement
with the exact result is only obtained for $\gamma _{2}\gtrsim \omega _{%
\mathrm{c}}$.

The inversion dynamics under the influence of phase relaxation can be
classified into different regimes. In the Rabi flopping regime, associated
with moderate dephasing, the inversion dynamics is governed by carrier-wave
Rabi oscillations. As discussed in Section 5, the steady state inversion
exhibits a significant phase sensitivity for sufficiently short and strong
pulses, or for pulses with phase dependent power spectra. For higher
dephasing rates, the near-transparency regime is reached, characterized by
an almost CE phase independent steady state inversion $w_{\mathrm{s}}\approx
0$. This regime can be associated with dephasing values of roughly $\gamma
_{2}/\omega _{\mathrm{c}}\approx 0.5\dots 5$ in Fig. \ref{cep6_4} and $%
\gamma _{2}/\omega _{\mathrm{c}}\approx 0.2\dots 5$ in Fig. \ref{cep6_5}. It
does not exist if the excitation is too weak, see Figs. \ref{cep6_2} and \ref%
{cep6_3}. In the limit of strong dephasing, $w_{\mathrm{s}}$ approaches its
initial value $w_{0}$, since the evolution of the inversion is suppressed.
The freezing of the quantum states can be interpreted as the quantum Zeno\
effect,\cite{mis77} where the decoherence of the states is induced by
frequent measurement or, more generally, by rapid system-environment
interactions. In the transition regime between transparency and freezing of
the states, the phase sensitivity slightly recovers for the rectangular
pulse, reaching an extremal value of $\Delta =-1.8\times 10^{-3}$ in the
inset of Fig. \ref{cep6_4}. In the quantum Zeno regime, corresponding to $%
\gamma _{2}/\omega _{\mathrm{c}}\gtrsim 5\times 10^{3}$ in Figs. \ref{cep6_4}
and \ref{cep6_5}, the modulation amplitude $\Delta $ and also the modulation
depth $\delta =\Delta /\left( \bar{w}_{\mathrm{s}}-w_{0}\right) $
asymptotically approach zero.

\section{Conclusion}

In conclusion, we have studied the carrier-envelope phase sensitivity of the
steady state inversion $w_{\mathrm{s}}$, remaining in a two-level system
after interaction with a pulse. We examined the invariance of $w_{\mathrm{s}%
} $ under time reversal of the field. Based on this property, we introduced
as parameters the phase averaged inversion $\bar{w}_{\mathrm{s}}$ and the
modulation amplitude $\Delta $. We discussed the two-level dynamics
analytically in the weak field limit, where a phase sensitivity of the
inversion could only be found for pulses with phase dependent power
spectrum. Beyond the linear regime, the phase sensitivity of $w_{\mathrm{s}}$
was studied numerically for excitation with rectangular and sinc-shaped
pulses. For the rectangular pulse, it was furthermore proven that the
modulation depth $\delta =\Delta /\left( \bar{w}_{\mathrm{s}}-w_{0}\right) $
does not depend on the pulse length, and an approximate analytical
expression for $\delta $ was derived applying almost degenerate perturbation
theory. The influence of phase relaxation on the inversion\ was investigated
based on the strong dephasing approximation and numerical simulations.

For interaction with few-cycle laser pulses, the two-level model predicts a
considerable phase dependence of $w_{\mathrm{s}}$ if the Rabi frequency
approaches or even exceeds the carrier frequency, see Fig. \ref{cep6_1}. The
necessary field strengths can be reached even for unamplified femtosecond
pulses directly out of the laser oscillator by tightly focusing the light
onto the sample.\cite{muc02} These model calculations show that the phase
dependent inversion is a promising route for the CE phase detection of
few-cycle laser pulses. One has to be aware of the fact that the
applicability of the two-level model to an atomic, molecular or solid state
system in strong fields has its limitations. Nevertheless, recent
theoretical and experimental work shows that basic features of the two-level
model are preserved for the strong field excitation with few-cycle laser
pulses.\cite{muc02,muc04}

The generated inversion can be detected optically, either by an additional
probe beam or by the short wavelength spectral component of the pulse
itself, which is delayed against the excitation part of the pulse and can
for example be chosen at resonance with the two-level system. It is also
possible to read out the excited electrons electronically with a field
ionization pulse.\cite{gri98} Experimentally, such a detector could be
realized by a gas or a metal vapor. For example, in the two-level
approximation, we obtain with $\Omega _{\mathrm{R}}=1.5\,\omega _{\mathrm{c}%
} $ a modulation amplitude of $\Delta =-0.34$ for interaction of two-cycle
Ti:sapphire laser pulses with a transition around $589\,\mathrm{nm}$,
corresponding to the famous D-lines in sodium. Another option might be the
use of artificial quantum systems like quantum dots.

\appendix

\renewcommand{\theequation}{A\arabic{equation}}
\setcounter{equation}{0}

\section*{Appendix A: Inversion invariance under time reversal of the field}

We now want to show that, for $\gamma _{1}=0$ and vanishing initial dipole
moment, $s_{1}=s_{2}=0$ for $t\rightarrow -\infty $, the remaining inversion
after passage of the pulse is invariant under time reversal of the field $%
\Omega \left( t\right) $. This should not be confused with the time-reversed
dynamics of the system. The Bloch equations in the rotating frame, Eq. (\ref%
{Bloch_rf}), can in matrix form be written as%
\begin{equation}
\mathbf{\dot{u}}=2\Omega M\mathbf{u-}\gamma _{2}\left( u_{1},u_{2},0\right)
^{\mathrm{T}}  \label{Bloch_rot}
\end{equation}%
with%
\begin{equation}
M\left( t\right) =\left( 
\begin{array}{ccc}
0 & 0 & \sin \left( \omega _{ba}t\right) \\ 
0 & 0 & \cos \left( \omega _{ba}t\right) \\ 
-\sin \left( \omega _{ba}t\right) & -\cos \left( \omega _{ba}t\right) & 0%
\end{array}%
\right)
\end{equation}%
and $\mathbf{u}=\left( u_{1},u_{2},u_{3}\right) ^{\mathrm{T}}$. We assume
that the pulse extends from $-t_{0}$ to $t_{0}$. As initial condition at $%
t=-t_{0}$, we choose $\mathbf{u}_{0}=\left( 0,0,w_{0}\right) ^{\mathrm{T}}$.
First we assume $\gamma _{2}=0$. The formal solution of Eq. (\ref{Bloch_rot}%
) is then given by%
\begin{eqnarray}
\mathbf{u}\left( t_{0}\right) &=&\hat{T}\exp \left(
\int_{-t_{0}}^{t_{0}}2\Omega \left( t\right) M\left( t\right) \mathrm{d}%
t\right) \mathbf{u}_{0}  \notag \\
&=&\lim_{N\rightarrow \infty }\prod_{n=-N}^{N}\exp \left[ 2\Omega \left(
t_{-n}\right) M_{-n}t_{0}/N\right] \mathbf{u}_{0}  \label{u(T)} \\
&=&\lim_{N\rightarrow \infty }\prod_{n=-N}^{N}\left[ I+2\Omega \left(
t_{-n}\right) M_{-n}t_{0}/N\right] \mathbf{u}_{0}  \notag
\end{eqnarray}%
with the time ordering operator $\hat{T}$, the unity matrix $I$, $%
t_{n}=t_{0}n/N$ and $M_{n}=M\left( t_{n}\right) $. By formally multiplying
out the last line of Eq. (\ref{u(T)}), we see that the solution is only
invariant under time reversal of an arbitrary field $\Omega \left( t\right) $
if 
\begin{equation}
M_{-n_{1}}M_{-n_{2}}\ldots M_{-n_{\ell }}\mathbf{u}_{0}=M_{n_{\ell }}\ldots
M_{n_{2}}M_{n_{1}}\mathbf{u}_{0}  \label{cond}
\end{equation}%
for any subset of indices $n$ with $-N\leq n_{1}<n_{2}<\ldots <n_{\ell }\leq
N$. For an even number $\ell $, we obtain%
\begin{eqnarray}
M_{n_{\ell }}\ldots M_{n_{2}}M_{n_{1}}\mathbf{u}_{0} &=&w_{0}\left(
-1\right) ^{\ell /2}  \notag \\
&&\times \prod_{m=1}^{\ell /2}\cos \left[ \omega _{ba}\left(
t_{n_{2m-1}}-t_{n_{2m}}\right) \right]  \notag \\
&&\times \left( 0,0,1\right) ^{\mathrm{T}}.
\end{eqnarray}%
If $\ell $ is odd, we obtain%
\begin{eqnarray}
M_{n_{\ell }}\ldots M_{n_{2}}M_{n_{1}}\mathbf{u}_{0} &=&w_{0}\left(
-1\right) ^{\left( \ell -1\right) /2}  \notag \\
&&\times \prod_{m=1}^{\left( \ell -1\right) /2}\cos \left[ \omega
_{ba}\left( t_{n_{2m-1}}-t_{n_{2m}}\right) \right]  \notag \\
&&\times \left( \sin\!\left( \omega _{ba}t_{\ell }\right) ,\cos\!\left( \omega
_{ba}t_{\ell }\right) ,0\right) ^{\mathrm{T}}.
\end{eqnarray}%
Thus, for odd $\ell $, Eq. (\ref{cond}) is only fulfilled for the third
vector component, and thus only $s_{3}\left( t_{0}\right) $ is invariant
under time reversal of the field.

The case with $\gamma _{2}\neq 0$ can be handled analogously by using the
transformation $u_{1}=\tilde{u}_{1}\exp \left( -\gamma _{2}t\right) $, $%
u_{2}=\tilde{u}_{2}\exp \left( -\gamma _{2}t\right) $. Eq. (\ref{cond}) must
now be fulfilled for the matrix 
\begin{widetext}
\begin{equation}
\tilde{M}\left( t\right) =\left( 
\begin{array}{ccc}
0 & 0 & \sin \left( \omega _{ba}t\right) \exp \left( \gamma _{2}t\right) \\ 
0 & 0 & \cos \left( \omega _{ba}t\right) \exp \left( \gamma _{2}t\right) \\ 
-\sin \left( \omega _{ba}t\right) \exp \left( -\gamma _{2}t\right) & -\cos
\left( \omega _{ba}t\right) \exp \left( -\gamma _{2}t\right) & 0%
\end{array}%
\right) .
\end{equation}%
\end{widetext}
For an even number $\ell $, we obtain%
\begin{eqnarray}
\tilde{M}_{n_{\ell }}\ldots \tilde{M}_{n_{2}}\tilde{M}_{n_{1}}\mathbf{u}_{0}
&=&w_{0}\left( -1\right) ^{\ell /2}  \notag \\
&&\times \prod_{m=1}^{\ell /2}\big\{\exp \left[ \gamma _{2}\left(
t_{n_{2m-1}}-t_{n_{2m}}\right) \right]  \notag \\
&&\times \cos \left[ \omega _{ba}\left( t_{n_{2m-1}}-t_{n_{2m}}\right) %
\right] \big\}  \notag \\
&&\times \left( 0,0,1\right) ^{\mathrm{T}}.
\end{eqnarray}%
If $\ell $ is odd, we obtain%
\begin{eqnarray}
\tilde{M}_{n_{\ell }}\ldots \tilde{M}_{n_{2}}\tilde{M}_{n_{1}}\mathbf{u}_{0}
&=&w_{0}\left( -1\right) ^{\left( \ell -1\right) /2}  \notag \\
&&\times \prod_{m=1}^{\left( \ell -1\right) /2}\big\{\exp \left[ \gamma
_{2}\left( t_{n_{2m-1}}-t_{n_{2m}}\right) \right]   \notag \\
&&\times \cos \left[ \omega _{ba}\left( t_{n_{2m-1}}-t_{n_{2m}}\right) %
\right] \big\}  \notag \\
&&\times \exp \left( \gamma _{2}t_{\ell }\right)   \notag \\
&&\times \left( \sin \left( \omega _{ba}t_{\ell }\right) ,\cos \left( \omega
_{ba}t_{\ell }\right) ,0\right) ^{\mathrm{T}}.
\end{eqnarray}%
Again, Eq. (\ref{cond}) is only fulfilled for the third vector component,
and thus only $s_{3}\left( t_{0}\right) $ is invariant under time reversal
of the field.

\renewcommand{\theequation}{B\arabic{equation}}
\setcounter{equation}{0}

\section*{Appendix B: Modulation depth for the rectangular pulse}

In the following, we want to show that for interaction with a rectangular
pulse shape, the modulation depth $\delta $, defined in Eq. (\ref{delta}),
does not depend on the pulse duration $T$. Assuming no dissipation, we can
describe the state of the two-level system by the ket $\left\vert \psi
\left( t\right) \right\rangle =c_{a}\left( t\right) \left\vert
a\right\rangle +c_{b}\left( t\right) \left\vert b\right\rangle $, with the
state vector $\mathbf{c}\left( t\right) \mathbf{=}\left( c_{a}\left(
t\right) ,c_{b}\left( t\right) \right) ^{\mathrm{T}}$. The initial condition 
$\rho _{ab}=\left( s_{1}+\mathrm{i}s_{2}\right) /2=c_{a}^{\ast }c_{b}=0$,
see Section 3, is here implemented by choosing the initial state vector $%
\mathbf{c}_{0}\mathbf{=}\left( 1,0\right) ^{\mathrm{T}}$, corresponding to $%
w_{0}=-1$. The evolution of the system can be described by the Schr\"{o}%
dinger equation. For an exciting field with period $T_{0}$, we can bring the
solution into the form%
\begin{equation*}
\mathbf{c}_{n}=U^{n}\mathbf{c}_{0},
\end{equation*}%
where $U$ is the unitary transformation matrix, and $\mathbf{c}_{n}$ is the
state vector after $n$ periods.\cite{gri98} We diagonalize $U$ by means of a
unitary matrix $T$, $\mathbf{c}_{n}^{\prime }=T\mathbf{c}_{n}$, $U^{\prime
}=TUT^{-1}$, and obtain for the $j$th component of $\mathbf{c}_{n}^{\prime }$%
\begin{equation*}
c_{nj}^{\prime }=\exp \left( -\mathrm{i}nW_{j}T_{0}/\hbar \right)
c_{0j}^{\prime }.
\end{equation*}%
Here, $W_{j}$ are the Floquet energies, which do not depend on the CE phase $%
\phi _{\mathrm{CE}}$ of the exciting field. By back transformation, we
obtain the occupation probability of the upper level%
\begin{equation}
\begin{array}{l}
\left| c_{b}\left( nT_{0}\right) \right| ^{2} \\ 
\;=\left| \frac{\left[ \exp \left( -\mathrm{i}nW_{b}T_{0}/\hbar \right)
-\exp \left( -\mathrm{i}nW_{a}T_{0}/\hbar \right) \right] T_{aa}T_{ba}}{%
T_{aa}T_{bb}-T_{ab}T_{ba}}\right| ^{2},%
\end{array}
\label{cb}
\end{equation}
and the inversion is given by $w\left( nT_{0}\right) =2\left\vert
c_{b}\left( nT_{0}\right) \right\vert ^{2}-1$. The matrix elements $T_{ij}$
depend on $\phi _{\mathrm{CE}}$, but not on the time $nT_{0}$. Thus, the
time dependence cancels out in the expression for the modulation depth, Eq.\
(\ref{delta}). We note that for the\ initial condition $s_{1}=s_{2}=0$, the
modulation depth is invariant with respect to $w_{0}$, since the choice of $%
w_{0}$ just leads to a scaling of the Bloch vector solution $\mathbf{s}%
\left( t\right) $ in Eq. (\ref{blo}). Thus, the proof is also valid for $%
w_{0}\neq -1$.

\renewcommand{\theequation}{C\arabic{equation}}
\setcounter{equation}{0}

\section*{Appendix C: Almost degenerate perturbation theory}

Following the formalism in Ref.\ 28, we can derive a perturbative
result for the phase dependent inversion. In this Appendix, the references
to equations refer to Ref.\ 28, unless otherwise stated, and we
adopt the nomenclature used there, with $\omega _{0}=\omega _{ba}$, $\omega
=\omega _{\mathrm{c}}$, and $\lambda =-\Omega _{\mathrm{R}}/2$. As initial
condition, we assume $w_{0}=-1$ as in Appendix B (also see remark at the end
of Appendix B), corresponding to $\rho _{\alpha \alpha }\left( t_{0}\right)
=1$, $\rho _{\beta \beta }\left( t_{0}\right) =\rho _{\alpha \beta }\left(
t_{0}\right) =\rho _{\beta \alpha }\left( t_{0}\right) =0$ in Ref.\ 28. From Eq. (1.33), we get after a few straightforward manipulations
the upper-state population $\rho _{\beta \beta }\left( t\right) =\left[
w\left( t\right) +1\right] /2$ in a two-level system%
\begin{eqnarray}
\rho _{\beta \beta }\!\left( t\right)  &=&\left| \sum_{m}\left\langle \beta
,m\right| \left.\!\chi _{+}\right\rangle e^{\mathrm{i}\omega mt}\right|
^{2}\left| \sum_{m}\left\langle \alpha ,m\right| \left.\! \chi
_{+}\right\rangle e^{\mathrm{i}\omega mt_{0}}\right| ^{2}  \notag \\
&&+\left| \sum_{m}\left\langle \beta ,m\right| \left.\! \chi _{-}\right\rangle
e^{\mathrm{i}\omega mt}\right| ^{2}\left| \sum_{m}\left\langle \alpha
,m\right| \left.\! \chi _{-}\right\rangle e^{\mathrm{i}\omega mt_{0}}\right|
^{2}  \notag \\
&&+2\Re \Bigg\{e^{2\mathrm{i}q\left( t-t_{0}\right) }\left(
\sum_{m}\left\langle \alpha ,m\right| \left.\! \chi _{+}\right\rangle e^{%
\mathrm{i}\omega mt_{0}}\right)   \notag \\
&&\times \left( \sum_{m}\left\langle \chi _{-}\right| \left.\! \alpha
,m\right\rangle e^{-\mathrm{i}\omega mt_{0}}\right)   \notag \\
&&\times \left( \sum_{m}\left\langle \chi _{+}\right| \left.\! \beta
,m\right\rangle e^{-\mathrm{i}\omega mt}\right)   \notag \\
&&\times \left( \sum_{m}\left\langle \beta ,m\right| \left.\! \chi
_{-}\right\rangle e^{\mathrm{i}\omega mt}\right) \Bigg\}.
\end{eqnarray}%
Here, $\left| \alpha ,m\right\rangle $, $\left| \beta ,m\right\rangle $ are
the pseudostates, and $\left| \chi _{+}\right\rangle $, $\left| \chi
_{-}\right\rangle $ are the Floquet modes of the two-level system. The
frequency $q$ is given in Eq. (1.71). With Eqs. (1.66), (1.73), (1.74),
(1.76), we obtain in first-order approximation 
\begin{eqnarray}
\left| \chi _{+}\right\rangle &=&N\big[\cos \left( \theta \right) \left(
\left| \alpha ,0\right\rangle -\eta \left| \beta ,1\right\rangle \right) 
\notag \\
&&+\sin \left( \theta \right) \left( \left| \beta ,-1\right\rangle +\eta
\left| \alpha ,-2\right\rangle \right) \big],  \notag \\
\left| \chi _{-}\right\rangle &=&N\big[-\sin \left( \theta \right) \left(
\left| \alpha ,0\right\rangle -\eta \left| \beta ,1\right\rangle \right) \\
&&+\cos \left( \theta \right) \left( \left| \beta ,-1\right\rangle +\eta
\left| \alpha ,-2\right\rangle \right) \big],  \notag
\end{eqnarray}%
with the normalized field $\eta =\lambda /\left( \omega +\omega _{0}\right) $%
. The normalization constant $N$ is given in Eq. (1.80b), and $\sin \left(
\theta \right) $, $\cos \left( \theta \right) $ are defined in Eq. (1.72).
Furthermore, by inserting Eq. (1.77) in Eq. (1.73), we obtain within the
first-order approximation%
\begin{eqnarray}
\epsilon _{jj} &=&-\hbar \omega _{0}/2-\eta ^{2}\hbar \left( \omega
_{0}+\omega \right) ,  \notag \\
\epsilon _{kk} &=&\hbar \omega _{0}/2-\hbar \omega +\eta ^{2}\hbar \left(
\omega _{0}+\omega \right) , \\
~\epsilon _{jk} &=&\epsilon _{kj}=\eta \hbar \left( \omega _{0}+\omega
\right) -\eta ^{3}\hbar \left( \omega _{0}+\omega \right) .  \notag
\end{eqnarray}%
The case $\phi _{\mathrm{CE}}=0$ corresponds to a cos pulse extending from $%
t_{0}=-T/2$ to $t=T/2$, where $T$ is equal to or an integer multiple of the
carrier period $T_{0}=2\pi /\omega $. We obtain for the remaining population
after passage of the pulse%
\begin{eqnarray}
\rho _{\beta \beta }\left( T/2\right) &=&2N^{4}\left[ \sin\!\left( \theta
\right) -\eta \cos \left( \theta \right) \right] ^{2}  \notag \\
&&\times \left[ \cos\!\left( \theta \right) +\eta \sin\!\left( \theta \right) %
\right] ^{2}\left[ 1-\cos\!\left( 2qT\right) \right] .
\end{eqnarray}%
The case $\phi _{\mathrm{CE}}=-\pi /2$ corresponds to a cos pulse extending
from $t_{0}=-T/2-T_{0}/4$ to $t=T/2-T_{0}/4$. We obtain%
\begin{eqnarray}
\rho _{\beta \beta }\left( T/2-T_{0}/4\right)  &=&2N^{4}\left[ \sin \left(
\theta \right) +\eta \cos \left( \theta \right) \right] ^{2}  \notag \\
&&\times \left[ \cos \left( \theta \right) -\eta \sin \left( \theta \right) %
\right] ^{2}  \notag \\
&&\times \left[ 1-\cos \left( 2qT\right) \right] .
\end{eqnarray}%
With Eq. (\ref{delta}) in our paper, we get the modulation depth%
\begin{equation}
\delta =\frac{2\eta \sin \left( \theta \right) \cos \left( \theta \right) %
\left[ \sin \left( \theta \right) ^{2}-\cos \left( \theta \right) ^{2}\right]
\left( 1-\eta ^{2}\right) }{\sin \left( \theta \right) ^{2}\cos \left(
\theta \right) ^{2}\left( 1-6\eta ^{2}+\eta ^{4}\right) +\eta ^{2}}.
\label{delta_p2}
\end{equation}%
We now insert Eq. (1.72), making use of the relations $\sin ^{2}\left(
\theta \right) -\cos ^{2}\left( \theta \right) =-\Delta /q$, $\sin \left(
\theta \right) \cos \left( \theta \right) =\epsilon _{jk}/\left( 2q\hbar
\right) $. The frequencies $q$ and $\Delta $ are given in Eq. (1.71). We
multiply the enumerator and denominator of our expression for $\delta $ by $%
q^{2}/\eta ^{2}$, and subsequently keep only the terms up to second order in 
$\eta $. Making the transition to the nomenclature used in our paper, with $%
\omega _{0}\rightarrow \omega _{ba}$, $\omega \rightarrow \omega _{\mathrm{c}%
}$, and $\eta \rightarrow -\Omega _{\mathrm{R}}/\left[ 2\left( \omega
_{ba}+\omega _{\mathrm{c}}\right) \right] $, we finally obtain the
perturbative result given in Eq. (\ref{delta_p}) in our paper.

\section*{Acknowledgment}

This study was supported by NSF under contract ECS-0217358, by contracts
ONR-N00014-02-1-0717 and AFOSR FA9550-04-1-0011, and by the Deutsche
Forschungsgemeinschaft under contract Mo 850/2-1. O.D.M. gratefully
acknowledges support by the Alexander von Humboldt Foundation. The work of
M.W. is supported by projects DFG-We 1497/11-1 and DFG-We 1497/9-1.

\end{document}